\documentclass[showpacs,aps,amsmath,amssymb,nofootinbib,notitlepage]{revtex4-1}
\usepackage[latin9]{inputenc}
\usepackage{amsmath}
\usepackage{graphicx}
\makeatletter
\usepackage{color}      
\usepackage{dcolumn}
\usepackage{bm}         
\usepackage{flushend}
\makeatother

\newcommand{\gev}{\text{GeV}}
\newcommand{\tev}{\text{TeV}}

\begin{document}

\title{Light Stops in a minimal $U(1)_x$ extension of the MSSM}

\author{R. M. Capdevilla, A. Delgado, A. Martin}

\affiliation{Physics Department, University of Notre Dame, Notre Dame, IN 46556, United States }

\begin{abstract}

In order to reproduce the measured mass of the Higgs boson $m_h=125\,\gev$ in the minimal supersymmetric standard model, one usually has to rely on heavy stops.
By introducing a new gauge sector, the Higgs mass gets a tree-level contribution via a non-decoupling $D$-term, and $m_h=125\,\gev$ can be obtained with lighter stops. In this paper, we study the values of the stops masses needed to achieve the correct Higgs mass in a setup where the gauge group is extended by a single $U (1)_x$  interaction.
We derive the experimental limits on the mass of the $Z'$ gauge boson in this setup, then discuss how the stops masses vary as a function of the free parameters introduced by the new sector.
We find that the correct Higgs mass can be reproduced with stops in a region between $700 - 800\,\gev$ and a $Z'$ resonance close to the $2.5\,\tev$ bound from the run-I of the LHC, or in a higher region $800 - 900$ GeV if the $Z'$ resonance is heavier (3.1 TeV). This region of parameter space will be quickly accessible at run-II of the LHC, and we discuss the impact of the projected run-II bounds on the $U(1)_x$ parameter space. We also discuss the phenomenology of the Higgs-like particles introduced to break $U(1)_x$ and conclude their effects are too small to be detected at current colliders.
\end{abstract}


\maketitle


\section{Introduction}

Supersymmetry is an elegant solution to the hierarchy problem, i.e. the difference between the Planck mass and the electroweak scale. In the Minimal Supersymmetric Standard Model (MSSM) the Higgs mass is not a free parameter, but it is given in terms of the  gauge couplings. In the decoupling limit, the tree-level mass turns out to be of the order of the $Z$ boson mass, in clear contradiction with the experimental results. This value is lifted by  loop corrections to the Higgs mass, which appear as contributions to the Higgs quartic coupling via virtual sparticles \cite{Casas:1994us,Carena:1995bx,Carena:1995wu,Haber:1996fp,Ambrosanio:2001xb,Carena:2002es}. For heavy enough sparticles, the loop corrections can push the mass of the Higgs boson up to $125\,\gev$~\cite{Draper:2011aa,Brummer:2012ns}; and as it is well known, the largest contribution to the radiatively corrected Higgs mass comes from stops loops.
In large areas of the parameter space the masses needed to reproduce the Higgs mass are above the LHC reach.
One way to allow light stops is by modifying the tree-level quartic of the Higgs boson via new $F$ or $D$ term contributions, so that one does not have to rely on large loop corrections to obtain a viable Higgs mass.

The effect of extended gauge groups on the Higgs mass was initially noticed in the framework of String Theory inspired $E_6$ Grand Unified groups \cite{Gunion:1986ky,Gunion:1987jd,Comelli:1992nu}. In particular, in \cite{Comelli:1992nu} the authors point out that the upper bounds on the Higgs mass can be raised up to 160 GeV. The LEP2 era placed the bound, $m_h > 115$ GeV (non-favorable for supersymmetric models), and in order to explain such heavy Higgs, the MSSM by itself requires large fine tuning. For this reason, D-terms extensions of the MSSM were proposed as a mechanism to raise the tree-level prediction and escape from the experimental limit without introducing fine tuning. A pioneering work in this research program can be found in \cite{Batra:2003nj}, and similar ideas in \cite{Maloney:2004rc,Martinez:2004rh,Babu:2004xg}. In this program, many possibilities were explored for the extra gauge groups (all these groups break down to the SM group)
$SU(2)_{1}\otimes SU(2)_{2}$ \cite{Batra:2003nj,Morrissey:2005uza,Medina:2009ey,Chiang:2009kb},
$SU(2)_{L}\otimes SU(2)_{R}\otimes U(1)_{B-L}$ \cite{Zhang:2008jm},
$SU(3)_{W}\otimes U(1)_{x}$ \cite{Bellazzini:2009ix} (with the differences of which fields are charged with respect to which groups),
a simple $U(1)$ \cite{Batra:2003nj,Lodone:2010kt},
and other more exotic groups \cite{Maloney:2004rc,Martinez:2004rh,Babu:2004xg,Cai:2012bsa}.
Even though the main motivation to include the non decoupled D-terms was to minimize the fine tuning, when the Higgs mass was still unknown, the parameter space was rather unconstrained.


In light of the Higgs discovery in 2012, there has been renewed interest in $D$-term extensions \cite{Endo:2011gy,An:2012vp,D'Agnolo:2012mj,Auzzi:2012dv,Cheung:2012zq,Craig:2012bs,Perez:2014gta,Dimopoulos:2014aua,Bertuzzo:2014sma,McGarrie:2014xxa,Tobioka:2015tpa,Belanger:2015cra}. In most of these references, the motivation for an extended gauge group is to obtain the Higgs mass in a natural supersymmetric scenario i.e to minimize the fine tuning.  Going into more detail, in Ref. \cite{McGarrie:2014xxa} the authors discuss in general terms the implications of including D-terms for what is called ``vector Higgs'' and ``chiral Higgs'' cases \cite{Craig:2012bs}. These cases have to do with the way in which the Higgs doublets are charged under the extra gauge group(s). Another example is Ref. \cite{Bertuzzo:2014sma}, where the authors discuss the fine tuning of the new gauge sector in a model with extra $SU(2)$ interactions and its relation with the uncertainty in the precision measurements of the Higgs-gluon-gluon and Higgs-gamma-gamma effective couplings. These authors find that a fine-tuning better than $20\%$ is compatible with a 125 GeV Higgs and no deviations in the Higgs effective couplings measurements. They also find the limits in stop masses $m_{\tilde{t}}\gtrsim300$ GeV based on these measurements and fine-tuning arguments, in agreement with previous work \cite{D'Agnolo:2012mj,Gupta:2012fy,Farina:2013ssa,Fan:2014txa}, but do not consider the combination of fine tuning and direct stop limits. Also, very recently, in Ref. \cite{Belanger:2015cra} the authors studied the contribution of the new sector to the dark matter relic abundance and they performed scans on the parameter space where the constrains coming from Higgs physics, collider searches for $Z'$ and sparticles, and flavor physics are satisfied.

In this paper we study the minimal $U(1)_x$ extension of the MSSM for its simplicity.
Given the value of the Higgs mass and the prediction from minimal supersymmetry, it is known that more than $30\%$ of the Higgs mass comes from loop correction. In the MSSM this means that the stops masses have to be at the multi-TeV scale, unless there is significant mixing in the stop sector, and therefore above the reach of the LHC. Current experimental bounds for stops explore supersymmetric simplified scenarios discarding stop masses around 400 - 700 GeV (depending on the mass of the lightest superpartner). If the lightest stop is detected close to the current bounds, the MSSM would demand a very big mixing term in order to explain the Higgs mass with such a light stop and if both stops are below 1 TeV it is fair to say that the MSSM would be unable to explain the Higgs mass independently of the value of the mixing. The goal of this paper is to explore an alternative supersymmetric scenario in which stops are allowed to be light without the need of large mixing because the Higgs mass in enhanced at tree level by non-decoupling D-terms. 
 We present an up-to-date analysis of the $U(1)$ parameter space, taking all recent and relevant LHC searches into account. That includes constraints on the $Z'$, stops, and Higgs mass. We point out where the lightest particles could be, and what that would mean for run-II. This paper is agnostic about the UV supersymmetry setup, so its a study of the IR parameter space.

The rest of this paper is organized as follow: Sec.~\ref{sec: Extra G. Interactions}
presents the model, the charges and the contribution of the D-term
to the Higgs mass. Section \ref{sec: Zprim-limits} shows the lower
limits for the mass of the $Z'$ gauge boson that we calculated for
our model and in comparison with the experimental bounds. We follow an analysis similar to the one presented in \cite{Safonov:2010qe,Arcadi:2013qia,Hayden:2013sra} and we compare the predicted cross section with the experimental constraints \cite{Chatrchyan:2012oaa,Chatrchyan:2013qha,Chatrchyan:2013fea,Aad:2014jra,Aad:2014aqa}.
 In Sec.~\ref{sec: Stops} we study different phenomenological aspects of the model: the Higgs mass as a function of the parameters of the new sector, a comparison between the supersymmetry breaking scale and the soft mass parameter $m_{\phi}$, the stop masses, and a discussion about the $\phi$ fields. We then present our conclusions.


\section{An Extra $U(1)_x$ Gauge Interaction \label{sec: Extra G. Interactions}}

Following~\cite{Batra:2003nj},  we extend the MSSM gauge group with a new $U(1)_x$ symmetry. In this model the MSSM superfields $\hat{L}$, $\hat{Q}$, $\hat{E}^c$, $\hat{D}^c$, $\hat{H}_u$, $\hat{U}^c$ and $\hat{H}_d$ are charged under the new gauge interaction as: 0, 0, 1/2, 1/2, 1/2, -1/2 and -1/2. In order to be anomaly free, the model introduces right-handed neutrinos $\hat{N}$ with the opposite charge of the right-handed leptons. The model also introduces a pair of fields $\hat{\phi}$ and $\hat{\phi^c}$ with charges $\pm1/2$ under $U(1)_x$.

The superpotential is slightly modified from the MSSM:
\begin{equation}
\hat{W}=\hat{W}_{MSSM}+\lambda\hat{S}(\hat{\phi}\hat{\phi}^{c}-\omega^{2}),\label{eq: W}
\end{equation}
where the $\hat\phi$ fields are responsible for breaking the extra gauge symmetry and the presence of the singlet $\hat{S}$ guarantees the existence of a supersymmetric minimum. 
The scalar potential consists of the usual MSSM contributions, plus three new pieces:
\begin{equation}
\Delta V_{soft}=m_{\phi}^{2}(|\phi|^{2}+|\phi^{c}|^{2})+m_{S}^{2}|S|^{2}+\lambda A_{\lambda}S(\phi\phi^{c}-\omega^{2})+h.c.\label{eq: Vsoft}
\end{equation}
\begin{equation}
\Delta V_{F}=\lambda^2|\phi\phi^c-\omega^{2}|^{2}+\lambda^{2}|S|^{2}(|\phi|^{2}+|\phi^{c}|^{2}),\label{eq: VF}
\end{equation}
\begin{equation}
\Delta V_{D}=\frac{g_{x}^{2}}{8}\left(|H_{u}|^{2}-|H_{d}|^{2}+|\phi|^{2}-|\phi^{c}|^{2}+...\right)^{2}.\label{eq: VD}
\end{equation}

If we suppose that EW breaking is a perturbation to the breaking of the $U(1)_x$, the 
vacuum of $V=V_{soft}+V_{F}+V_{D}$ is where $S$ does not get a vacuum expectation value (vev)
but $\phi$ and $\phi^{c}$ do. The difference of these vevs is proportional to the Higgs vev and therefore can be neglected. Integrating out the $\phi_{i}$ fields, the MSSM scalar potential $V$ gets an extra contribution \cite{Batra:2003nj} that depends on the mass of the $Z'$ boson $M_{Z'}=g_{x}v_{\phi}$, $v_{\phi} = \langle \phi \rangle \cong \langle \phi^c \rangle$ and the soft mass parameter $m_{\phi}$. Including this new contribution, the tree-level Higgs mass is given by (in the decoupling limit $m_{A}\gg M_{Z}$)
\begin{equation}
m_{h_{\phi}}^{2}=\frac{1}{2}\left[g^{2}+g'^{2}+g_{x}^{2}\left(1+\frac{M_{Z'}^{2}}{2m_{\phi}^{2}}\right)^{-1}\right]v^{2}\cos^{2}(2\beta),
\label{eq: m_h_tree}
\end{equation}
where $g$ and $g'$ are the couplings of the $SU(2)_L$ and $U(1)_Y$ gauge groups, $\tan \beta$ is the ratio of the two  MSSM higgses and $v=174$  GeV is the vev of the Higgs boson. If supersymmetry were not broken, there would be no effect on the Higgs potential because the $\langle \phi \rangle$ and $\langle \phi^c \rangle $ pieces in the D-term would exactly cancel. However, when supersymmetry is softly broken, this generates an effect on the Higgs mass\footnote{This fact can be seen by setting $m_{\phi}=0$ in equation (\ref{eq: m_h_tree}), which shows that with no supersymmetry breaking there is no effect on the Higgs mass.}. By including the $U(1)_x$ $D$-term contribution, the Higgs quartic coupling is increased, and $m_h = 125\, \gev$ can be achieved with lighter stops. In fact, in the presence of the $D$ term piece, stops that are too heavy will cause the Higgs mass to overshoot the experimental value. This model has some immediate consequences: the appearance of a new Z boson-like resonance,  relatively light stops, and two new Higgs-like $\phi_i$ particles. All these signatures of the model can be studied using current and near future data from the LHC. In the next sections we explore the relation between the experimental bounds for a $Z'$ resonance and light stops for a 8 TeV and a 14 TeV collider. 


\section{Limits on the $Z'$ mass\label{sec: Zprim-limits}}

The $Z'$ in this model couples to SM fermions both through direct couplings and through $Z-Z'$ mixing. The expressions for the different couplings and mixing angle depend on the ratio $M_Z^2/M_{Z'}^2$ and can be found in \cite{Agashe:2014kda,Langacker:2008yv}. The decay of the $Z'$ is dominated by the channels $Z'\rightarrow f\bar{f}$, $Z'\rightarrow W^{+}W^{-}$ and $Z'\rightarrow Z\:h$. The partial widths in all modes grow linearly with $M_{Z'}$, so the branching ratios are basically constants in the limit when $M_{Z'}\gg M_Z$.

\begin{figure*}[h!]   
\includegraphics[scale=0.475]{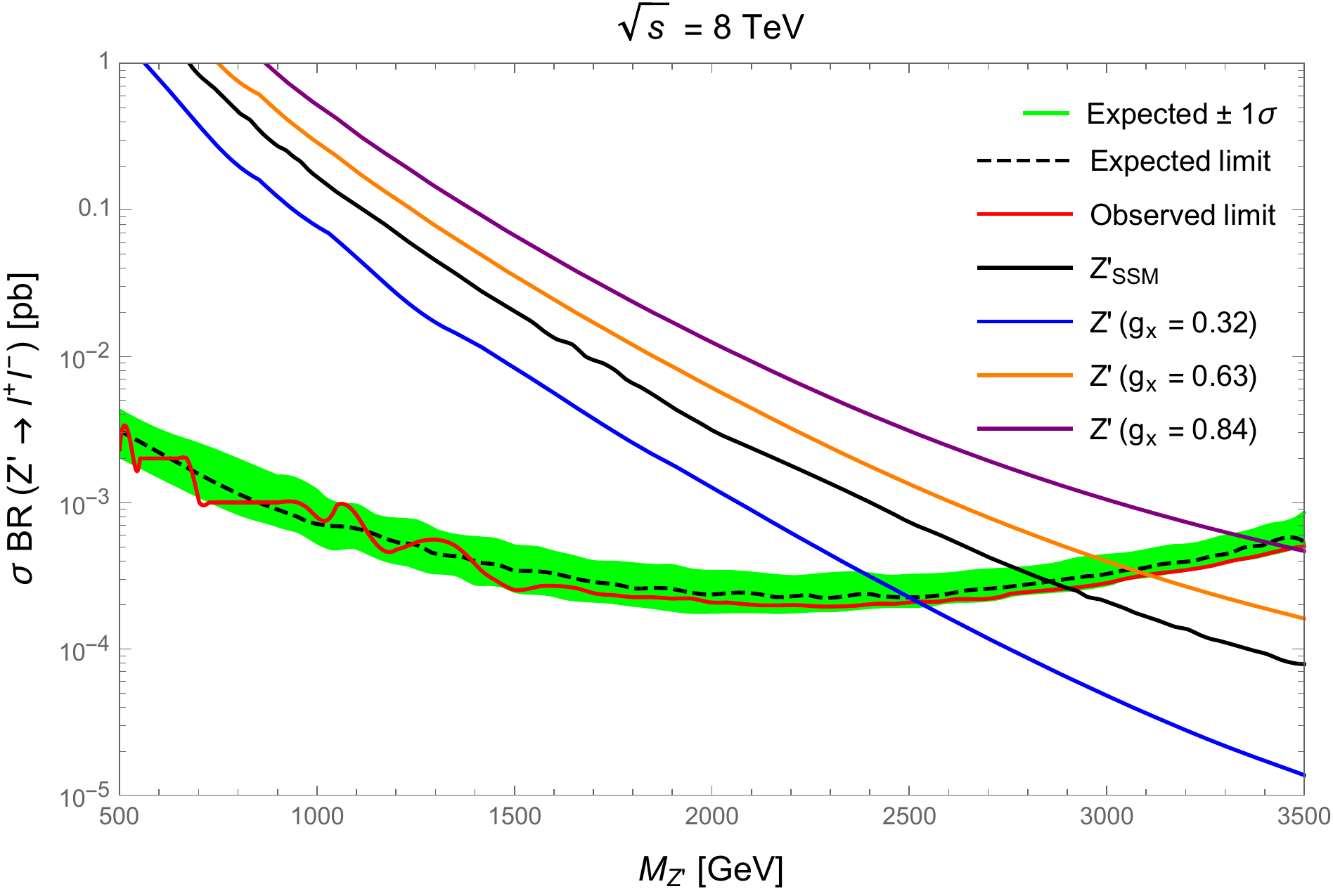}\includegraphics[scale=0.495]{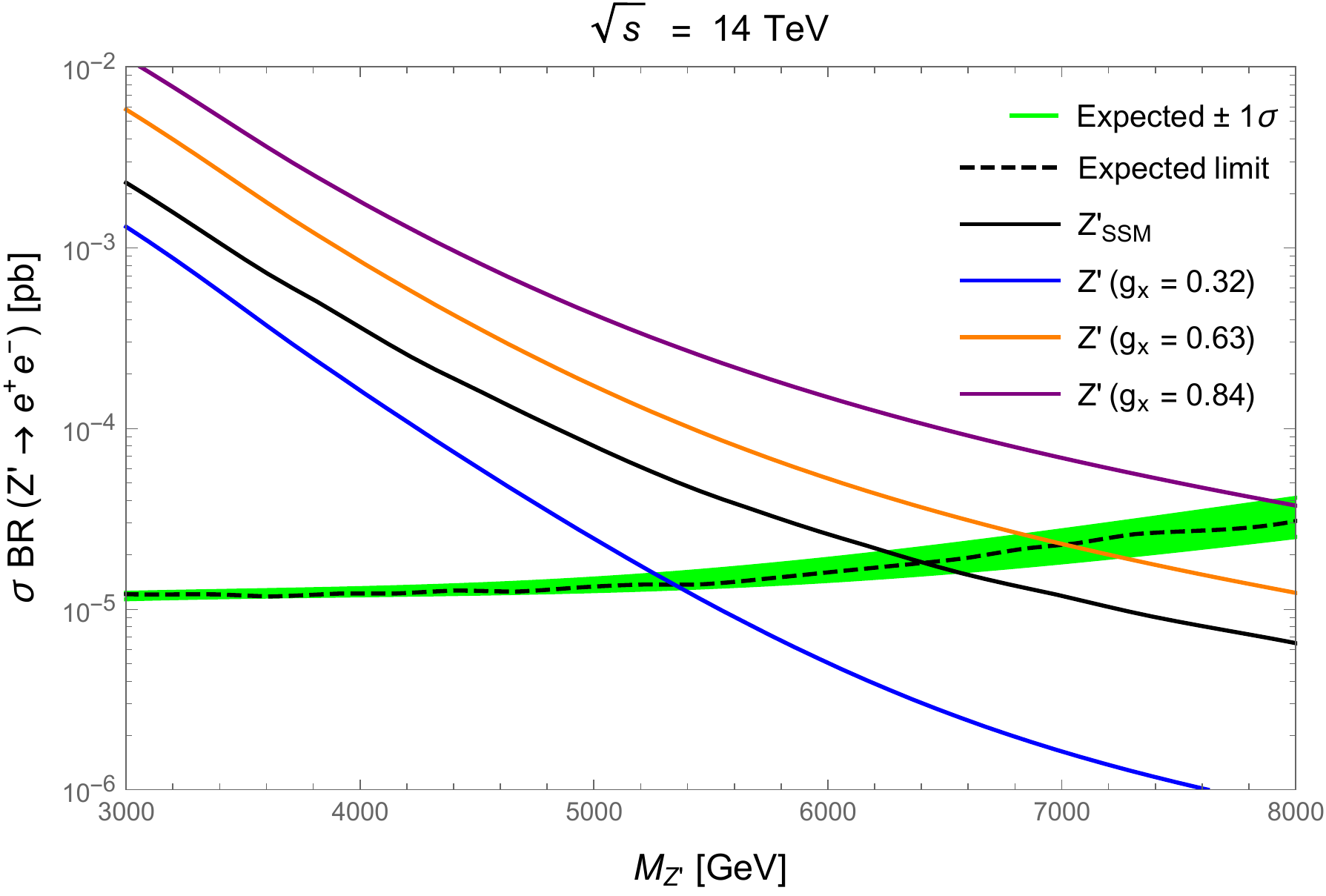}
\protect\protect\caption{\label{fig: sigma Z'-ll}Left: Total cross section times branching ratio of the $Z'$ in the di-lepton channel  as a function of the $Z'$ boson mass. The interception of the blue, orange and purple lines and the observed limit represent the lower bounds for the $Z'$ mass of our model for the given values of the gauge coupling $g_{x}$. The black line shows the limit on a Sequential Standard Model (SSM) $Z'$~\cite{Altarelli:1989ff}, for comparison. Right: The 14 TeV prediction for the cross section of the present model in a $e^+e^-$ final state channel, in comparison with the expected limit after 300 fb$^{-1}$ of integrated luminosity \cite{Hayden:2013sra}.} 
\end{figure*}

 In this paper we perform a similar analysis to \cite{Safonov:2010qe,Arcadi:2013qia,Hayden:2013sra} and set limits on new gauge bosons using their decays to dileptons. For a given $Z'$ mass and $U(1)_x$ coupling, we compare the cross section for production ($\sigma$) times branching ratio ($BR$) into two leptons with experimental data.  The left-hand panel of Fig. \ref{fig: sigma Z'-ll} presents such a comparison. The intersection of the observed limit (red line in Fig.~\ref{fig: sigma Z'-ll}) and the theory line for a particular value of $g_x$ gives the current lower $Z'$ mass bounds. For $U(1)_x$ gauge couplings of $g_x = 0.32,\,0.63$ and $0.84$, we find the current limits as of run-I of the LHC are $M_{Z'}\gtrsim2.5,\,3.1$ and 3.5 TeV, respectively. Since the $Z'$ can decay to superparticles, we have to assume a spectrum in order to derive the limits. Specifically, to generate Fig.~\ref{fig: sigma Z'-ll} we took the following naturalness-inspired spectrum: $\mu = 300\, \gev,\, M_{stops}=1$ TeV, tan$\beta=20$, and all other superpartner masses at $4$ TeV. With this setup, the only superpartners the $Z'$ can decay to are third generation squarks and Higgsinos, and we find the $Z'$ limits are fairly insensitive to the masses of these states. Notice how the $g_x=0.63$ exclusion is similar to the limits on the Sequential Standard Model (SSM) \cite{Altarelli:1989ff}. This is expected as this coupling is about the same size of the SM $g$ coupling and because the supersymmetric channels do not contribute considerable to the width. 

The right-hand panel of figure  \ref{fig: sigma Z'-ll}  shows the same analysis repeated for a $14\, \tev$ LHC and assuming 300 fb$^{-1}$ of luminosity. For the expected limits, we use the projections in Ref.~\cite{Hayden:2013sra}. We find the lower bounds increase to $M_{Z'}\ge 5.4,\,7.0$ and $8.2\,\tev$ for $U(1)_x$ couplings $g_x = 0.32,\,0.63$ and 0.84. These increases can be compared with the SSM, where we find the limit raises from 2.9 TeV to 6.5 TeV. In what follows, we will both explore regions of the parameter space close to the lower limits for $M_Z'$ coming from the 8 TeV run of the LHC and regions within the reach of run II of  LHC.


\section{Phenomenology\label{sec: Stops}}

\subsection*{Higgs Mass}

In the present model, the radiatively corrected Higgs mass in the decoupling limit can be written as
\begin{equation}
m_h^{2}=M_{Z}^{2} \cos^{2}(2\beta)+ \frac{1}{2}g_{x}^{2}v^{2}\cos^{2}(2\beta)\left(1+\frac{M_{Z'}^{2}}{2m_{\phi}^{2}}\right)^{-1}+\Delta m^2_{\rm loops},
\label{eq: m_h_loops2}
\end{equation}
where the loop contributions depend (among other supersymmetric and soft parameters) on the stop mixing parameter $X_t = A_t - \mu \cot \beta$ and the supersymmetry breaking scale defined as
\begin{equation}
M_{SUSY}^{2}=m_{\tilde{t}_{1}}m_{\tilde{t}_{2}}.
\label{eq: Msusy}
\end{equation}

In order to estimate the Higgs mass we have used SUSYHD
\cite{Vega:2015fna}, a recent public code that calculates
the MSSM Higgs mass including two-loop threshold corrections to the
quartic Standard Model (SM) Higgs coupling. The code runs
down the SM parameters using three-loop renormalization group equations with a leading four-loop
QCD contribution to the strong gauge coupling. To calculate the Higgs mass, we have defined the following parameters: degenerate squarks ($m_{D_3}$ plus first and second generations) and slepton masses $m_{\tilde{q}}=m_{\tilde{l}} =4$ TeV, equal Wino and Bino masses $M_{2}=M_{1}=2$ TeV, gluino mass $m_{\tilde{g}}=1.4$ TeV (close to the lower bounds reported by the LHC \cite{Aad:2015lea}) and a supersymmetric Higgsino mass of $\mu=300$ GeV. Working with this spectrum, in order to reproduce the Higgs mass in the {\em pure} MSSM, we need $M_{SUSY}$ to be around 1.6 TeV and 11.5 TeV in the cases of max-mixing ($\hat{X_t} = X_t / M_{SUSY} = \sqrt6$) and non-mixing ($\hat{X_t} = 0$), respectively\footnote{These extremum values of mixing maximize and minimize the loop contribution from stops to the Higgs mass.}. This behavior is consistent with previous work using different codes \cite{Bagnaschi:2014rsa,Draper:2013oza}.  Increasing the gaugino masses to 4 TeV causes a variation of less than 20\% in $M_{SUSY}$.

\begin{figure*}[h!]   
\includegraphics[scale=0.545]{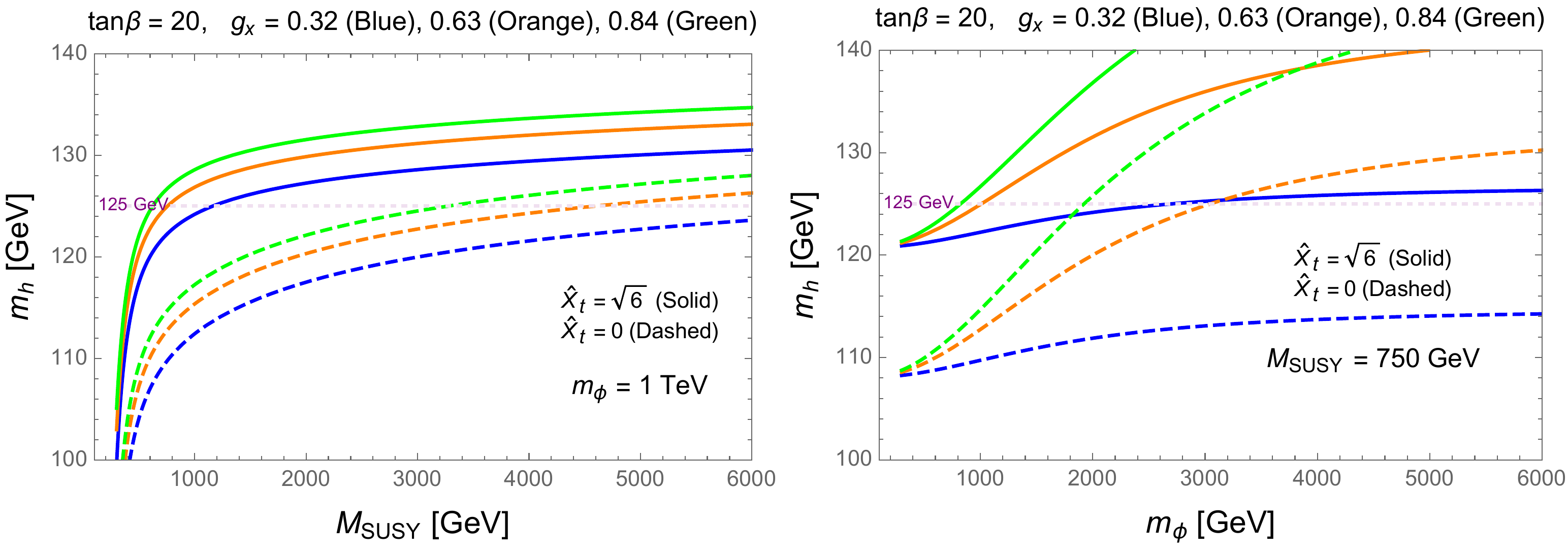}
\protect\protect\caption{\label{fig: Mh} Left (Right): Higgs mass as a function of $M_{SUSY}$ ($m_{\phi}$) for different values of the coupling $g_x$ and fixing $m_{\phi}=1$ TeV ($M_{SUSY} = 700$ GeV). The solid lines correspond to the Max-Mixing case and the dashed lines to the Non-Mixing case. In both panels we have used the lower bound values for $M_{Z'}$ that correspond to the given $g_x$ values.}
\end{figure*}

Moving to the $U(1)_x$ extension, the picture changes, and the values of the Higgs mass are shown in Fig.~\ref{fig: Mh} for different values of the $U(1)_x$ coupling.
The left panel shows the Higgs mass as a function of $M_{SUSY}$ with $m_{\phi}$ fixed to 1 TeV, while the right panel shows the Higgs mass for variable $m_{\phi}$ and fixed $M_{SUSY}=750$ GeV (close to the highest lower bound reported for the stop mass \cite{Aad:2015pfx,Chatrchyan:2013xna,Khachatryan:2015wza,Khachatryan:2014doa,Chatrchyan:2013mya}). Focusing on the left panel, for $m_{\phi}=$ 1 TeV with $g_x = 0.32$, the new $D$-term gives a contribution to the Higgs mass such that the value of $M_{SUSY}$ reduces from 11.5 TeV to 8.1 TeV and from 1.6 TeV to 1.1 TeV for the no-mixing and max mixing cases, respectively.
For larger couplings we see how the correct Higgs mass is achieved for even smaller values of $M_{SUSY}$, for example $g_x=0.63$ implies $M_{SUSY}=0.8$ TeV (max-mix) and $M_{SUSY}=4.6$ TeV (non-mix), while $g_x=0.84$ implies $M_{SUSY}=0.6$ TeV (max-mix) and $M_{SUSY}=3.3$ TeV (non-mix).  Moving to the right panel, we see that if $M_{SUSY}$ is fixed to $750\,\gev$, we cannot generate $m_h = 125$ GeV in the no-mixing scenario and $g_x = 0.32$. For intermediate coupling, e.g. $g_x = 0.63$, the no-mixing scenario can be compatible with $m_h = 125\, \gev$ for $m_{\phi}$ around 3 TeV. For the max-mixing cases, small couplings $g_x=0.32$ and $m_{\phi}$ around 2.6 TeV suffice.
 
 Rather than picking  $m_{\phi}$ or $M_{SUSY}$ and calculating the Higgs mass as a function of those two variables, we can enforce $m_h = 125\,\gev$ and show the relationship between the two scales. This is shown below in Fig.~\ref{fig: Msusy vs mphi} for the same set of $U(1)_x$ couplings and stop mixing assumptions as in Fig. \ref{fig: Mh}.  When the D-term contribution is zero ($m_{\phi}=0$), the value of $M_{SUSY}$ is equal to the aforementioned MSSM critical values. As the $D$-term piece increases, the necessary value of $M_{SUSY}$ drops.
 In Fig.~\ref{fig: Msusy vs mphi}, we are varying the $\phi, \phi^c$ soft mass and $U(1)_x$ coupling for fixed $M_{Z'}$ (equivalent to keeping the $U(1)_x$ vev fixed). At the left panel, for a coupling of $g_x = 0.32,\,0.63$ and 0.84, we use the (current) corresponding lower bound for the $Z'$ mass, i.e. $M_{Z'} = 2.5$, 3.1 and 3.5 TeV, respectively. In the right panel, we use only two values of the coupling, as it can be inferred what would happen for an intermediate one. We then increase the $Z'$ mass by 1 TeV from the lower bound given by the 8 TeV search, from 3.5 to 4.5 TeV. This analysis shows what would happen if the new search for resonances in dilepton channels finds a $Z'$ boson just above the exclusions bounds set by run I of the LHC. 
 
 From Fig.~\ref{fig: Msusy vs mphi} alone, it looks like larger $m_{\phi}$ is always better since it increases the $D$-term piece. This is not correct for a couple of reasons. First, the expression in Eq.~(\ref{eq: m_h_loops2}) came from integrating out $\phi, \phi^c$ at tree level and neglecting any running effects between $m_{\phi}$ and the weak scale. For very large $m_{\phi}$, the effects of running cannot be ignored. Second, the $\phi, \phi^c$ scalars are connected to the Higgs through the $U(1)_x$ D-term and therefore feed into the radiative corrections for the Higgs soft masses ($\delta m^2_{H_u} \sim g^2_x\, m^2_{\phi}$). So increasing the soft parameter $m_{\phi}$ increases the tree-level Higgs mass but also the loop correction. This effect is not significant at the scale range we explored in this paper.
 

\begin{figure*}[h!]   
\includegraphics[scale=0.545]{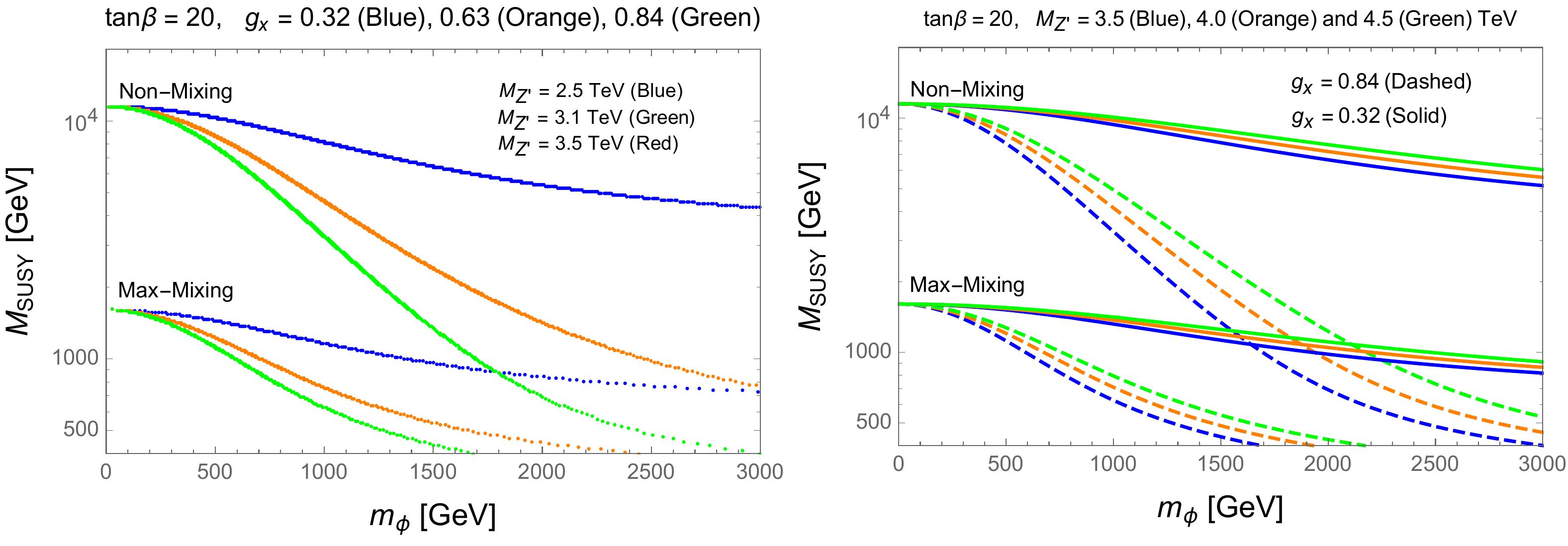}
\protect\protect\caption{\label{fig: Msusy vs mphi}Contours in the plane ($M_{SUSY}, m_{\phi}$) for which the Higgs mass has the right value. We are using large $\tan\beta$ and, in the left panel we have three values of the new coupling $g_x=0.32,\,0.63$ and 0.84 paired with the lower bounds for $M_{Z'}$ that correspond to those couplings. In the right panel we have used the range $M_{Z'} = 3.5 - 4.5$ TeV for two values of the new coupling. In both panels, the higher contours correspond to the non-mixing case and the lower ones to the max-mixing case.}
\end{figure*}

\subsection*{Stops Masses: 8 TeV Analysis \label{sec: Stops2}}

\begin{figure*}[h!]   
\includegraphics[scale=0.565]{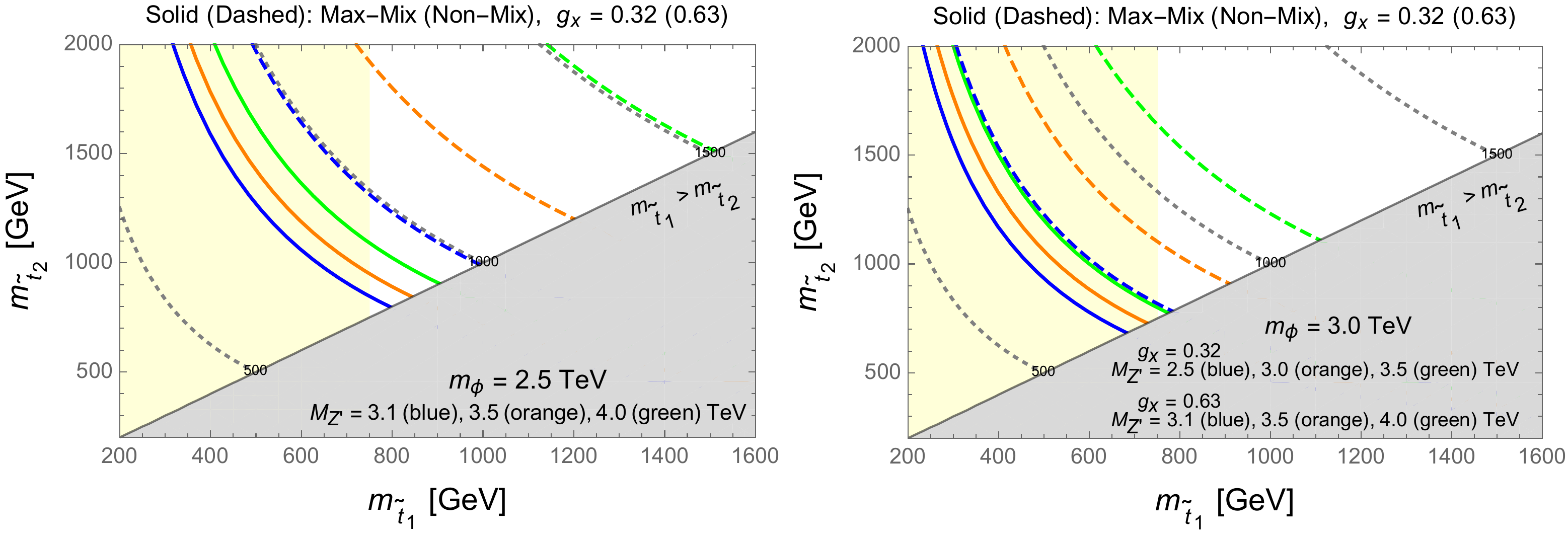}
\protect\protect\caption{\label{fig: Stops} Masses of the stop eigenstates for large $\tan\beta$, max-mixing with a small coupling $g_x=0.32$ (solid lines) and non-mixing with a larger coupling $g_x=0.63$ (dashed lines) cases. The gray-shaded region are excluded because $m_{\tilde{t}_{2}}>m_{\tilde{t}_{1}}$ and the gray-dotted contours correspond to fixed values of $M_{SUSY}$. The yellow region correspond to a conservative bound that excludes the lightest stop for $m_{\tilde{t}_1}<750$ GeV (refer to the text for details).}
\end{figure*}

Using the definition (\ref{eq: Msusy}) we can find the relation between the masses of the stop eigenstates corresponding to the contours showed in Fig.~\ref{fig: Msusy vs mphi}.
With this, we can see how the individual stops masses -- relevant for all direct LHC searches --  depend on the new parameters $M_{Z'}$, $m_{\phi}$ and $g_x$. This relationship is shown by the solid and dashed contours in Fig.~\ref{fig: Stops} below for a fixed $m_{\phi}$. 
A given set of $U(1)_x$ parameters defines $M_{SUSY}$ and appears as a line in Fig.~\ref{fig: Stops}. For example, for $g_x=0.32$ and max-mixing, $m_{\phi}=2.5$ TeV and the values $M_{Z'}=3.1$, 3.5 and 4.0 TeV we generate the solid contours in the left panel of Fig.~ \ref{fig: Stops} for which $M_{SUSY}=$ 830, 880 and 940 GeV, respectively. Increasing the gauge coupling to $g_x=0.63$, using the same $m_{\phi}$ and $M_{Z'}$ and assuming no stop mixing yields the dashed contours; the corresponding $M_{SUSY}$ values are 980, 1200 and 1500 GeV, respectively. We perform a similar analysis in the right panel of Fig.~\ref{fig: Stops} but consider a wider range of $M_{Z'}$ and $m_{\phi}$. Both panels show the same trend;  increasing the gauge coupling or the soft term $m_{\phi}$ increases the D-term contribution, thereby causing the value of $M_{SUSY}$ to decrease and shifting the stop mass contours to lower values. This behavior can be seen by comparing the solid orange line in the left panel with the green solid line in the right panel; these lines correspond to $g_x=0.32$, max-mixing, $M_{Z'}=3.5$ TeV and $m_{\phi}=2.5$ and 3.0 TeV, respectively. On the other hand, as we increase the mass of the $Z'$, the effects of the D-terms reduce, and $M_{SUSY}$ and the stop mass contours move to higher values.  For example, increasing $M_{Z'}$ from 2.5 to 3.5 TeV changes $M_{SUSY}$ from 730 to 810 GeV, for $g_x=0.32$ and max-mixing, and increasing $M_{Z'}$ from 3.1 to 4.0 TeV, $M_{SUSY}$ changes from 760 to 1100 GeV for $g_x=0.63$ with non-mixing.


Next, we discuss the experimental bounds on stops. Although there are a plethora of stop searches at the LHC, each involves some degree of assumption about the stop branching ratio, the mass difference between the stop and its decay products, and the composition of the stops and lightest superparticle (LSP) \cite{Aad:2015pfx,Chatrchyan:2013xna,Khachatryan:2015wza,Khachatryan:2014doa,Chatrchyan:2013mya}. As a result, it is almost impossible to derive a model-independent bound. In the supersymmetric spectrum we have used so far the LSP mass is 300 GeV, value for which there is currently {\em no} bound on the lightest stop\footnote{Obviously, the mass of the lightest stop has to  be always greater than $\mu$.}. Decreasing $\mu$ (and therefore the LSP) mass will lead to an limit between roughly 400 GeV to 750 GeV for $\tilde{t}_1$. In Fig. \ref{fig: Stops} we show the 750 GeV limit as the lower bound for the lightest stop, although this should be viewed as a conservative bound -- most stops searches suppose a 100\% BR of the stop into top plus neutralino, while in our case there is a sensible probability of stop decay into bottom quark plus chargino. In either decay mode, the bounds become looser if the spectrum is highly compressed since the stop decay products become soft and cannot be efficiently identified. As these bounds are independent on the details of the $U(1)_x$ sector, they will not change if we increase or decrease the mass of the $Z'$.


Clearly there is still a large set of reasonable  $U(1)_x$ parameters where the stops evade the current bounds.
For maximally mixed stops, $g_x = 0.32, m_{\phi} = 2.5\, \tev$ and $M_{Z'} = 3.1\, \tev$, we achieve a $125\, \gev$ Higgs with the stops nearly degenerate and lying just beyond the strictest run I bound.
As the $Z'$ mass is increased with the other $U(1)_x$ quantities fixed, the stop mass contours move farther away from the run I limits. While this implies heavier stops
it also allows the stops to be more widely separated. For instance, upping $M_{Z'}$ to 4 TeV, the lightest stop can still sit just beyond the run I bound if the heavier stop is pushed to $1.3\, \tev$. For the no-mixing scenario, the stop mass contours are farther from the exclusion bands.
For $g_x = 0.63,\, m_{\phi} = 2.5\, \tev$ and $M_{Z'} = 3.5\, \tev$, we can have $M_{SUSY} \sim 1\, \tev$; 
 in this case, the stops can either be degenerate and sit well beyond the current bound, or split, with $m_{\tilde t_1} = 750\, \gev, m_{\tilde t_2} \sim 1.5\, \tev$. As the most natural regions occur when the stops -- either one or both -- lie just beyond the current limit, the next run of the LHC will quickly rule in or out these possibilities. In the next section, we give a brief analysis of how the parameter space will change in the scenario where no hints of the $U(1)_x$ model are uncovered at run II.  


\subsection*{A 14 TeV Analysis \label{sec: 14tev}}

The $U(1)_x$ parameter space will be impacted by run II LHC searches for the stops (both $\tilde t_1$ and $\tilde t_2$), and for the $Z'$.

Let us first consider the 14 TeV lower limits on the $Z'$, which were shown in section \ref{sec: Zprim-limits}. After $300\, \text{fb}^{-1}$ of integrated luminosity, the values of $M_{Z'}$ increase by a factor of $\sim2.3$ for the range of couplings presented here. If the $Z'$ mass is increased keeping all other $U(1)_x$ parameters fixed, the $D$-term contribution shrinks and must compensated by a larger $M_{SUSY}$ (stop loop). The  factor of $\sim 2.3$ increase in $M_{Z'}$ corresponds to an increase in  $M_{SUSY}$ by a factor of 1.7 for maximally-mixed stops and a factor of 3.9 for unmixed stops.
This increase would move the contours for the stops showed in Fig.~\ref{fig: Stops} to regions beyond 2 TeV for either stop mixing scenario. In order to balance this effect, we can consider heavier $m_{\phi}$. We find that to reproduce the same contours presented in Fig.~\ref{fig: Stops} it is necessary to increase $m_{\phi}$ by a factor of 2.3. However, as mentioned earlier, $m_{\phi}$ feeds back into the Higgs soft masses, so heavier $m_{\phi}$ might not be accommodated.

Turning to searches for supersymmetry at the 14 TeV LHC, if the gluino is not found, the new lower bounds for its mass are expected to be around 2 TeV \cite{ATL-PHYS-PUB-2015-005}. With this variation, the values of $M_{SUSY}$ presented here will increase by around 60 GeV for max-mixing, but they will not change for non-mixing, indicating that the gluino does not have a significant impact on the analysis here presented. Repeating the exercise for the stops, the expected exclusion limit at run II is $m_{\tilde t_1} \gtrsim 1\,\tev$\cite{ATL-PHYS-PUB-2015-005}. For stops consistent with this bound, the $D$ term contribution has to diminish, either by taking smaller coupling, smaller $m_{\phi}$, or larger $M_{Z'}$, From Fig.~\ref{fig: Stops}, we see that $1\, \tev$ stops are consistent with $g_x = 0.63,\ M_{Z'} = 4\, \tev,m_{\phi} = 3\, \tev$, or $M_{Z'} = 3.5\, \tev$, $m_{\phi} = 2.5\, \tev$. Both of these examples assume no stop mixing, since $1\, \tev$ maximally mixed stops are almost sufficient for $m_h = 125\, \gev$.
 
 Finally, at  run II of the LHC there is also a potential signal coming from the heaviest stop decaying into the lightest one; these searches are not yet competitive but could give another handle once the data at higher energy gets analyzed.

\subsection*{The Higgs-like $\phi$ particles \label{sec: phi}}

Another possible signature of this model comes from the extended scalar sector needed to break the $U(1)_x$. As can be seen from the fact that the $U(1)_x$ breaking occurs in a D-flat direction, the mixing between the mass eignenstate $\phi$ fields and the Higgs is highly suppressed. Therefore, single productions of $\phi$ has a very small cross-section and double production as a on-shell decay of the $Z'$ will be equally small. Therefore any possible signals coming from the new sector are very unlikely to have enough significance over the background.



\section*{Summary and Conclusions}

We have studied the interplay of an extra $U(1)_{x}$ non-decoupling D-term contribution to the Higgs quartic and the stop masses needed to reproduce $m_h=125$ GeV. The fact that the quartic coupling of the Higgs gets an extra tree-level contribution will, in general, mean that in order to reproduce $m_h=125$ GeV the value of the stop masses will have to be smaller than in the MSSM
We have analyzed the different regions of the parameter space of the model varying the mass of the new $Z'$ gauge boson $M_{Z'}$, the new gauge coupling $g_{x}$, and the soft mass term $m_{\phi}$ of the fields $\phi_i$ that are introduced to break the extra gauge symmetry. As we increase the magnitude of the coupling or the value of the soft parameter, the contribution from the D-term to the Higgs mass increases, therefore the contribution from the stops is lower and their masses decrease.  In this scenario we found that a coupling of $g_{x}=0.63$ and $m_{\phi}=3$ TeV would allow the lightest stop to be in a region lower than 1 TeV for the non-mixing case, this should be compared to the value of 11.5 TeV needed for the stops in the MSSM  with the same mixing. For the max-mixing scenario the MSSM loop contributions require stops around 1.6 TeV in order to reproduce the Higgs mass. In the $U(1)_x$ model with a coupling of $g_{x}=0.32$ and $M_{Z'}=2.5$ TeV, we find the stops can be in a region around 700 - 800 GeV; if the $Z'$ mass is bigger (3.1 TeV), the stops can be heavier 800 - 900 GeV. In general, as we increase the mass of the $Z'$ gauge boson, the contribution from the D-term becomes less significant.

If the 14 TeV searches for supersymmetry do not find the stops, considering the predicted exclusion regions, we found that the lower bounds for a $Z'$ resonance are pushed up considerable by a factor of 2.3. This causes the contribution from D-terms to the Higgs mass to be suppressed which means that the stops have to be heavier.
This effect can be balanced by increasing the soft parameter $m_{\phi}$, however radiative corrections on the Higgs soft mass parameter $m_{H_u}$ indicate that $m_{\phi}$ cannot be arbitrarily large.

We finally discussed the extra Higgs-like $\phi$ particles. These fields mix with the other MSSM CP-even Higgs particles, so their presence can modify the production rate of the MSSM Higgs boson. We found that for the region of parameter space explored in this paper the effects of the new particles are too small to be detected in the experimental data.

A general conclusion for models where the Higgs mass is a prediction, like in supersymmetry, is that the measured value of 125 GeV can put both lower and \emph{upper} bounds on the masses of new particles.

\section*{Acknowledgments}

This work was partially supported by the National Science Foundation under Grants
No. PHY-1215979 and No. PHY-1417118.
\bibliography{MSSMpU1Bib}

\end{document}